\documentclass[twocolumn,prl]{revtex4}

\usepackage{amsmath,amssymb,bm}
\usepackage{graphicx}
\usepackage{hyperref}
\usepackage{epstopdf}

\begin{document}

\title{Efimov states in Li-Cs mixtures within a minimal model}

\author{N. T. Zinner}
\affiliation{Department of Physics and Astronomy - Aarhus University, Ny
Munkegade, bygn. 1520, DK-8000 {\AA}rhus C, Denmark}
\author{N. G. Nygaard}
\affiliation{Lundbeck Foundation Theoretical Center for Quantum System Research,
Department of Physics and Astronomy, Aarhus University, DK-8000 Aarhus C, Denmark}

\begin{abstract}
We use a minimal zero-range model for describing the 
bound state spectrum of three-body states consisting of two Cesium and 
one Lithium atom. Using a broad Feshbach resonance model for the two-body
interactions, we show that recent experimental data 
can be described surprisingly well for particular values of the three-body 
parameter that governs the short-range behavior of the atomic potentials
and is outside the scope of the zero-range model. Studying the spectrum 
as a function of the three-body parameter suggests that the lowest state
seen in experiment could be influenced by finite range corrections. 
We also consider the question of Fermi degeneracy and corresponding 
Pauli blocking of the Lithium atoms on the Efimov states.
\end{abstract}
\maketitle

\paragraph{Introduction.} 
The ability to study few-body states in the low-energy regime with cold
atoms has allowed the observation of Efimov three-body states \cite{efi1970,kraemer2006}
and also resonances associated with bound states of four or more particles
\cite{ottenstein2008, pollack2009, zaccanti2009, gross2009, huckans2009,
williams2009, lompe2010a, gross2010, lompe2010b, nakajima2010,
nakajima2011, berninger2011, wild2012, machtey2012a, machtey2012b, knoop2012, zenesini2013, rem2013, hulet2013, roy2013}. 
The key property of these low-energy bound states is
a discrete scaling symmetry \cite{nie01,bra2006} that has been difficult to 
observe in experiments (a very recent paper presents evidence for the 
observation of a second Efimov resonance in Cesium \cite{huang2014}). 
Theory shows that systems with a large mass imbalance
between the bound state constituents are highly suited for studying the discrete
scaling since the scaling factor becomes small and more states should be 
observable within experimental constraints.

This search has prompted different groups to pursue cold atomic experiments with
mixtures of different atoms. In particular, recent experiments by 
the Weidem{\"u}ller group in Heidelberg \cite{repp2013} and the Chin group
in Chicago \cite{tung2013}
have pursued mixtures of $^6$Li and $^{133}$Cs which has a suitably low 
scaling factor of around 4.8 (depending on the number of resonant two-body
subsystems \cite{nie01}). The first experimental results have now been 
presented by both the Chicago group \cite{tung2014} and most recently by
the Heidelberg group \cite{pires2014}. Here we will focus mostly on 
the data presented in Ref.~\cite{tung2014} and make a comparison at 
the end.
In Ref.~\cite{tung2014} they find three consective
Efimov peaks that are attributed to the Cs-Cs-Li system 
when working with a Feshbach resonance located at a magnetic field of 842.75 Gauss (G). 
Since the Cs-Cs scattering length at these fields is negative there are no 
molecular thresholds to worry about and thus the system is ideal for studying
the formation of Efimov trimers from the three-atom continuum. The observations
are done by analyzing atom loss peaks associated with the threshold. 
In this short paper, we consider a minimal model for describing the 
experimental data that uses zero-range interactions and the simplest 
model for describing the two-body Feshbach resonances. Our results are in 
good agreement with the experiments although we do see some differences
for the most bound trimer that may be associated with finite range corrections.
This system could be useful for studying the effects of a many-body 
background on the few-body physics \cite{zinner2014}. We therefore 
consider the influence of the Fermi degeneracy in the Lithium component 
on the Efimov states by introducing a Pauli blocking in the light constituent.
For Lithium densities in the range $10^6$ to $10^{12}$ cm$^{-3}$ we find 
virtually no changes in the three-body thresholds for Efimov trimer formation.

\paragraph{Model.} 
To solve the Schr{\"o}dinger equation for the three-body problem with 
short-range interactions, 
we use a zero-range momentum-space method similar to 
the one introduced by Skorniakov and Ter-Martirosian \cite{STM} but 
suitably regularized by the method of Danilov \cite{danilov} (see the 
discussion in \cite{pri}). This regularization method has typically 
been applied to equal mass systems but it is straighforward to generalize
to our case with different masses (see the appendix of Ref.~\cite{nygaard2014}).
In the regularization we assume that all three two-body subsystems
are resonantly interacting in order to obtain the Efimov scale factor.
While this is not the case for the experimental Cs-Cs-Li system, the
Cs-Cs interaction is nevertheless very large and we expect the difference
to be small (in any case the difference in the Efimov scale factor is 
small, see Fig.~1 in Ref.~\cite{yamashita2013}).

The zero-range interactions are parameterized
by the two-body scattering lengths and in the general case of non-identical 
masses and scattering lengths this yields a set of coupled integral equations
that can be solved by discretization (see the appendix of Ref.~\cite{nygaard2014} 
for details on the formalism). It is important to note that we take
the scattering length of all subsystems and particularly its 
magnetic field dependence (see below) into account in our model, both
that of the two heavy-light and the heavy-heavy systems.
A necessity of a zero-range model is a 
three-body parameter, $\kappa$, that is used in the regularization of the 
equations. It can be physically interpreted as a short-range repulsive force
that is due to the hard-core nature of the inter-atomic potentials at short distance.
Here we assume that $\kappa$ is a constant and will not depend on any external 
magnetic fields.
We will work at zero temperature and without
any explicit decay channels in the model. The latter implies that we are 
making the assumption that there is a one-to-one correspondence between
the threshold at which an Efimov state appears out of the three-atom 
continuum and the loss peak in the measurement. This is a simplied picture
yet it is our express goal to explore the capabilities of a minimal model 
for describing the data. 

To model the two-body interactions that are tuned by Feshbach resonances
in experiments, we note that the resonances in the Cs-Cs-Li case are all
entrance channel dominated and of large resonance strength (also called 
broad resonances in many places). 
Therefore we use the simplest model where the scattering length has
the parameterization \cite{chin2010}
\begin{align}
a(B)=a_{bg}\left(1+\frac{\Delta}{B-B_0}\right).
\end{align}
Here $a_{bg}$ is the background scattering length away from the resonance, 
$\Delta$ is the resonance width, and $B_0$ is the position. The Cs-Cs 
system has a very wide resonance around 787 G \cite{ferlaino2011}. 
The Efimov states occur for fields around 840-850 G and in this regime
the Cs-Cs systems has a large negative scattering length. We will therefore
keep the parameters of this heavy-heavy subsystem fixed in our model 
with $B_0=787$ G, $\Delta=87.5$ G, and $a_{bg}=-1940a_0$, where $a_0$
is the Bohr radius. We have checked that changes in $\Delta$ and $a_{bg}$
for the Cs-Cs channel within the ranges that are given in the literature
\cite{ferlaino2011,lee2007,chin2010,berninger2013} have no noticable effect on the results 
discussed here. This is connected to the fact that this resonance is very
broad. On the other hand, for the Li-Cs subsystems we work with two parameter
sets below to reflect the different values given in recent papers \cite{repp2013,tung2013}.

\begin{figure}[ht!]
\centering
\includegraphics[scale=0.45,clip=true]{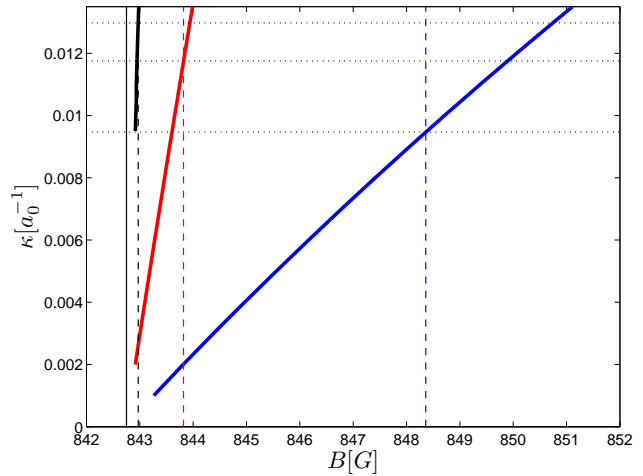}
\caption{Magnetic field positions at which Efimov trimers cross the three-atom 
continuum as function of the three-body parameter, $\kappa$. The solid
lines are the three resonances (they disappear for small but different $\kappa$ values).
The vertical dashed lines are the experimental values reported in Ref.~\cite{tung2014}. The
three horizontal dotted lines mark the positions at which the vertical dashed experimental 
results and solid theoretical calculation results cross (see the text). 
The vertical solid line shows the position 
of the Li-Cs Feshbach resonance.}
\label{fig1}
\end{figure}

\paragraph{Results.} 
We first consider the Li-Cs Feshbach resonance data given in Ref.~\cite{tung2013} 
(and updated in Ref.~\cite{tung2014}). The reported best fit value for the position 
is $B_0=842.75$ G, while the backgtround scattering length is $a_{bg}=-22a_0$. 
We use $\Delta=62$ G from Ref.~\cite{tung2013}. This specifies everything 
in our model except for the three-body parameter, $\kappa$, which we can then
vary. In Fig.~\ref{fig1} we show the calculated magnetic field positions 
as a function of $\kappa$ along with the experimental data. The thick solid
lines denote the position at which the three-body Efimov trimer hits the 
three-atom continuum. Each line terminates at a given value of $\kappa$ as
the spectrum is pushed upwards by decreasing $\kappa$ and thus less trimers
are present. To avoid confusion we stress that these theoretical lines have
been obtained by taking a fixed value of $\kappa$, then computing the spectrum 
at the three-atom continuum threshold, then changing to a new value of $\kappa$
and repeating this. These threshold spectra may then be connected into three 
solid lines as given in the figure. It is very important to notice that there 
is {\it no} magnetic field dependence of $\kappa$ in any of the calculations 
presented here, $\kappa$ is always a constant input parameter. 

The three dotted lines in Fig.~\ref{fig1} mark the points at which the vertical 
dashed experimental lines and the solid theoretical calculations overlap, i.e. 
it gives the value of $\kappa$ for which the minimal model reproduces exactly 
one of the experimental data points. Ideally, these three dotted lines would 
coincide implying that a single constant value of the three-body parameter $\kappa$
could reproduce the data in the minimal model. This is, however, not the case. 
We may turn the argument around and say that in a minimal model one needs to fix
$\kappa$ to an experimental data point in order for all parameters to be determined. 
This could be done for either of the three experimental points, but would of course
imply that the predictions of the model would be slightly off for the other two 
points. However, the values of the dotted lines lie in the range from $\kappa=0.0095a_{0}^{-1}$ 
to $\kappa=0.013a_{0}^{-1}$ so the variation is only around 25\% which we consider
very good for this minimal model.

We typically associate $\kappa$ 
with short-range physics that does not change from state to state. 
On the other
hand, the deeper Efimov states are more susceptible to finite-range corrections
\cite{sorensen2013}. 
This can cause the discrepancy and the difference in where the experimental 
results and theoretical calculations overlap for the three loss features.
The difference in $\kappa$ between the two states at the smaller magnetic fields 
(closer to the Feshbach resonance) is seen to be smaller as expected. A 
possible solution to this problem is to postulate a
variation
in $\kappa$ with magnetic field to describe experimental data. This has been noted 
also for equal mass Efimov studies in three-component $^6$Li systems and interpreted
as an energy-dependence of $\kappa$ \cite{nakajima2011}. Finite-range corrections can 
be one cause of such an effect. In a model where one also includes the effective 
range, $r_e$, the threshold for the most bound Efimov trimer will move
depending on the sign of $r_e$ \cite{sorensen2013}. The trimer line on the far
right in Fig.~\ref{fig1} would move to smaller magnetic field values for $r_e>0$
according to the results of Ref.~\cite{sorensen2013}. However, other models have 
found different behaviour \cite{schmidt2012} and a more precise model
such as a multi-channel framework \cite{nygaard2006} seems necessary to explore this 
issue. We stress again that we have used a constant $\kappa$ in all the calculations
reported here.

We may provide a slightly more quantitative estimate of the finite-range corrections
by using the results of Ref.~\cite{sorensen2013}. This can be done by considering the
relation between the scattering length at which an Efimov state emerges out of 
the three-atom continuum, $a_{-}^{n}$, and the energy on resonance of the same state,
$E_{\infty}^{n}$. One typically defines a momentum scale, $\kappa^{n}$, 
through $E_{\infty}^{n}=-\tfrac{\hbar^2(\kappa^{n})^{2}}{2m_r}$, where $m_r$ is the 
reduced mass. In Ref.~\cite{sorensen2013} the product of $\kappa^{0}a_{-}^{0}$ for the
ground state of equal mass systems was explored when including the effective range 
using different scattering models (see Fig.~5 of Ref.~\cite{sorensen2013}). To get
a simple estimate we now make some assumptions to relate our mass imbalanced case to 
the equal mass results. 

We will assume that the Li-Cs scattering length is the
important interaction and use the Feshbach resonance values given above to convert
between the magnetic field values and $a_-$. Second, we will with the three-body 
parameter $\kappa=0.01175a_{0}^{-1}$, a value for which the two excited Efimov 
states we calculate fit well with experimental magnetic field values according to 
Fig.~\ref{fig1}. Now we will use the second excited Efimov state as our reference 
point as it should be least sensitive to numerical details (this is one reason for 
the near insensitivity to $\kappa$ of this second excited state seen in Fig.~\ref{fig1}).
We may then compute $\kappa^{2}$ by numerically determining the energy of the second
excited state on resonance at $B=842.75$ G. We find a theoretical value of 
$\kappa^{2}a_{-}^{2}=-29.17$ (using simple units where $m_r=m_\textrm{Li}$). 
In a pure zero-range approach, we should then be able to get $\kappa^{0}$ from $\kappa^{2}$
by use of the scale factor, i.e. $\kappa^{0}=e^{2\pi/s_0}\kappa^{2}$, where 
$e^{\pi/s_0}\sim 4.8$ is the Efimov scaling factor for the Li-Cs system.
If we now use the experimental value of $a_{-}^{0}\sim -265.14a_{0}$ with this calculated 
value of $\kappa^{0}$ we find $\kappa^{0}a_{-}^{0}=-25.69$, which is about 12\%
lower than for the second excited state. From the equal mass case in Ref.~\cite{sorensen2013}
a lower values implies a negative value of $r_{e}/a_{bg}$, which is consistent with 
$r_e>0$ since $a_{bg}<0$ here. For an equal mass system a 12\% decrease corresponds to 
$r_e/a_{bg}$ between $-1$ and $-3$ roughly (see Fig.~5 in Ref.~\cite{sorensen2013}). 
If $\kappa^{0}a_{-}^{0}$ depends only weakly on the mass ratio in the system then 
we can estimate that a finite-range correction in the range $r_e\sim 22-66 a_{0}$ is
necessary to explain the deviation of the ground state Efimov state from the zero-range
model. Interestingly, the van der Waals length of Li-Cs is in this range $r_\textrm{vdW}=45a_0$
\cite{pires2014}. This could still be a coincidence as $r_e$ and $r_\textrm{vdW}$ are
generally different \cite{chin2010}. At this point we could not locate a theoretical 
value for $r_e$ in this system for comparison. If there is a stronger dependence
of finite-range corrections on the mass ratio then this will change the numbers 
found here. Taking these corrections into account in the momentum-space formalism
we have used in this work is not straightforward and is left for future studies.

Overall, we would describe the agreement as very
good given the simplicity of the model.
Moreover, it is interesting to note that
our zero temperature model so closely reproduces the data for the two states closest
to the Feshbach resonance. Also, the experiment finds very little change with temperature
of the Efimov resonance closest to resonance which should be the most sensitive one. 
A zero temperature formalism seems therefore to work well.

We have also studied a second set of parameters that are taken from Ref.~\cite{repp2013}. 
The resonance position is slightly changed to $B_0=843.5$ G, while $\Delta=60.4$ G
and $a_{bg}=-28.5a_0$. We do this to test the overall features and their robustness
while still keeping close contact with experimentally realistic numbers. However, since
the qualitative picture is identical to Fig.~\ref{fig1} we omit a figure. The 
only change is quantitative as the Efimov features move with the Feshbach resonance position.
This implies that the scenario seen in both experiments is quite robust and 
does not depend much on the specifics of the two-body resonances. We conclude 
that the Cs-Cs-Li system is extremely well-suited for the study of few-body physics.
In comparison to the most recent data from the Heidelberg group \cite{pires2014}, 
we note that the position of the Feshbach resonance reported there is now 
consistent with the Chicago value \cite{tung2014}. There is some discrepancy
in the value of the magnetic field of the first Efimov loss peak with a 
best fit value that is larger than in Ref.~\cite{tung2014}. The two other
features reported in the data have best fit values that are very close in 
both experiments. However, while the Heidelberg group also find evidence
of a third peak close to the Feshbach resonance, they cannot conclude 
that it displays the universal Efimov scaling behaviour. 

Recent data \cite{berninger2011} and a number of theoretical works 
\cite{naidon2011,chin2011,wang2012,sorensen2012,schmidt2012,naidon2012,ywang2012,naidon2014}
have suggested a new type of universality in the Efimov three-body problem 
where the three-body parameter, $\kappa$, is related to the two-body 
van der Waals length of the inter-atomic potential in a simple manner. 
In Fig.~\ref{fig1}, we see that a value of around $\kappa\sim 0.01a_{0}^{-1}$ 
is needed to reproduce the experimental data in our minimal model. Translating
this momentum scale into length yields $L=\kappa^{-1}=100a_0$ which is
almost exactly equal to the van der Waals length of the Cs-Cs system ($101a_0$).
This can be understood from a Born-Oppenheimer point of view where the 
light Li atom creates an effective potential for the two heavy Cs-Cs 
atoms where in the bound states can be calculated. The natural cut-off
on the universal part in this picture is the short-range Cs-Cs van der Waals
length. Here we have used a momentum space approach so a more precise translation
from $\kappa$ to length scale is necessary to put this on a more firm footing.

\begin{figure}[ht!]
\centering
\includegraphics[scale=0.45,clip=true]{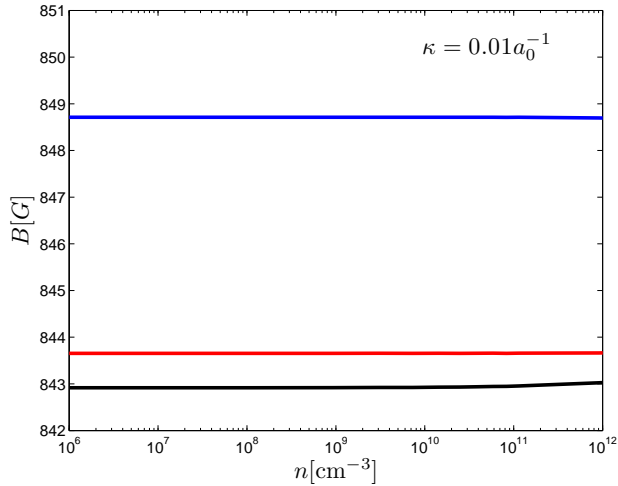}
\caption{Magnetic field positions of the Efimov trimers for 
$\kappa=0.01a_{0}^{-1}$ as function of the density of fermionic 
$^6$Li atoms, $n$. The horizontal axis is logarithmic and expends
over six decades.}
\label{fig2}
\end{figure}

Finally, we consider the question of whether the system could be suitable
for investigating the effects of quantum degeneracy on the three-body
physics. More specifically, we ask whether a finite density of fermionic
$^6$Li atoms can influence the loss peaks. We still assume zero temperature
as above. The effects of a Fermi sea on bound state complexes have been 
the subject of a lot of recent work \cite{nygaard2014,nishida2009,macneill2011,mathy2011,niemann2012,endo2013}.

We will now briefly explain our approach to including a Fermi sea (full details
can be found in the appendix of Ref.~\cite{nygaard2014}). 
Here we consider the simplest approximation where the Fermi sea is considered
inert, and thus its effect is Pauli blocking of states for the Lithium
atom constituent in the trimer. 
The momentum-space
three-body equations can be derived by considering the scattering of a single 
atom on a dimer (see Ref.~\cite{bra2006} for an extensive elaboration on how 
this is done). The bound state equations for three different atoms will 
generally yields a set of three coupled integral equations \cite{nygaard2014}, 
where the ingredients are the Green's function (propagator) for a single 
atom and the progator for a pair of atoms, the dimer propagator. To 
include a background Fermi sea in one of the components, one must 
do two things. First, in the propagator for the atom with the Fermi
sea one introduces a Pauli blokcing factor, i.e. $\theta(k_F-k)$ where
$\theta(x)$ is the Heaviside step function and $k_F$ is the Fermi momentum 
which is given by the density of the fermionic atom. This means that only states
with momenta above the Fermi sea are allowed. Likewise, when the atom 
with the Fermi sea is in a dimer one needs to modify the dimer propagator
by taking Pauli blocking of one of its constituents into account. This 
requires some calculation, details can be found in Ref.~\cite{nygaard2014}. 
One these two effects are takne into account, one may now solve the 
coupled integral equations for a given $k_F$ to obtain the bound states
and the spectral flow as a function of $k_F$. In this simple approach 
we do not consider particle-hole pairs in the Fermi sea. Such corrections
are expected bring some quantitative changes to thresholds but not to the
overall physical picture of a spectral flow. Within the Born-Oppenheimer
approximation, this was discussed in Ref.~\cite{macneill2011}
(see also the appendix to Ref.~\cite{nygaard2014} for a discussion).

We again make the assumption that the 
loss peak can be mapped to the three-atom continuum threshold position 
which will generally shift due to the Fermi sea \cite{nygaard2014}. In 
Fig.~\ref{fig2} we show the calculated loss peak positions as a 
function of the density of $^6$Li atoms. The three-body parameter
was taken to be $\kappa=0.01a_{0}^{-1}$. We have checked that the 
results in Fig.~\ref{fig2} are not sensitive to $\kappa$ as long
as it is chosen in the range where three states are found as in Fig.~\ref{fig1}
($\kappa>0.0095a_{0}^{-1}$). The density in Fig.~\ref{fig2} extends
over six decades from $10^6$ cm$^{-3}$ to $10^{12}$ cm$^{-3}$ which 
is an experimentally relevant range. In spite of this large
variation in density we find basically no effect on the thresholds.
There is a small upturn on the third Efimov resonance closest to the 
Feshbach resonance for the largest densities. However, given the 
uncertainties in the numerical procedure we cannot conclude that
this is a physical effect. These data suggest that studying the effect of 
Fermi degeneracy on Efimov physics could be difficult for this 
system. This fact is presumably connected to the fact that the 
mass imbalance in the system is very large. It is known \cite{nie01,bra2006}
that the effective three-body hyperradial equation has an attractive
inverse sqaure potential whose strength will increase with increasing
mass imbalance and thus make the resulting three-body states more 
bound. In turn, it requires a larger Fermi energy to perturb the 
states and thus we need much higher density of fermionic atoms to 
see a spectral flow.

\paragraph{Conclusions.}
Using a minimal zero-range three-body bound state calculation we have 
managed to reproduce experimental data on Efimov states in highly 
mass imbalanced systems quite well. The difference between the experimental 
data and the model was largest for the deepest bound Efimov state seen in 
experiment, yet it was still in fairly good agreement with the calculations. 
The lowest state is expected to have the largest finite-range corrections in 
a given spectrum and we speculate that this could be the cause. This is outside
the scope of our minimal model. We also estimate the effect of the fermionic
nature of the Lithium atoms by introducing a Pauli blocking effect. However,
in a large range of experimentally relevant densities we find virtually
no influence on the Efimov thresholds.

The current study does not take the decay of Efimov states into account. 
This can be done in different ways \cite{esry1999,bra2006,optical2013},
but requires an input parameter to describe the decay which goes into 
deep bound two-body states as there are no Feshbach molecules on this side
of the Feshbach resonance. 
Temperature is another effect that should be taken into account, particularly
on the width of the Efimov resonances as discussed in Ref.~\cite{tung2014}.
Furthermore, the potential effects of a condensate in the heavy Cesium 
component would be interesting to study. While the fermionic nature of 
the light component did not look important based on the current results,
a condensate could still have an interesting effect. For instance, 
the linear dispersion relation of (interacting) condensed bosons 
can cause changes in three-body dissociation thresholds as the condensate coherence length 
changes. This can be shown most clearly using the Born-Oppenheimer approximation for the 
case where the light particle is condensed \cite{zinner2013}. For the current
system it is the heavy component that could form a condensate and a different
approach is needed.

\end{document}